\documentclass[12pt,titlepage]{article}
\usepackage{amssymb}

\begin{document}

\title{Relativistic confinement of neutral fermions \\with a trigonometric
tangent potential}
\date{}
\author{Luis B. Castro\thanks{%
E-mail address: benito@feg.unesp.br (L.B. Castro)} and Antonio S. de Castro%
\thanks{%
E-mail address: castro@pesquisador.cnpq.br (A.S. de Castro)} \\
\\
UNESP - Campus de Guaratinguet\'{a}\\
Departamento de F\'{\i}sica e Qu\'{\i}mica\\
Av. Dr. Ariberto P. da Cunha, 333\\
12516-410 Guaratinguet\'{a} SP - Brasil }
\maketitle

\begin{abstract}
The problem of neutral fermions subject to a pseud\-oscalar potential is
investigated. Apart from the solutions for $E=\pm mc^{2}$, the problem is
mapped into the Sturm-Liouville equation. The case of a singular
trigonometric tangent potential ($\sim \mathrm{tan}\,\gamma x$) is exactly
solved and the complete set of solutions is discussed in some detail. It is
revealed that this intrinsically relativistic and true confining potential
is able to localize fermions into a region of space arbitrarily small
without the menace of particle-antiparticle production.
\end{abstract}

\section{Introduction}

The four-dimensional Dirac equation with a mixture of spherically symmetric
scalar, vector and anomalous magnetic-like (tensor) interactions can be
reduced to the two-dimensional Dirac equation with a mixture of scalar,
vector and pseudoscalar couplings when the fermion is limited to move in
just one direction $(p_{y}=p_{z}=0)$ \cite{str}. In this restricted motion
the scalar and vector interactions preserve their Lorentz structures while
the anomalous magnetic-like interaction becomes a pseudoscalar. This kind of
dimensional reduction does not necessarily imply that the reduced Dirac
equation describes a fermion in an unrealistic two-dimensional world. As a
matter of \ fact, since there is no spin flip in a one-dimensional motion
the two-dimensional version of \ the Dirac equation can be thought as that
one describing a fermion embedded in a four-dimensional space-time with
either spin up or spin down \cite{gre}. This happens because the
four-dimensional Dirac equation with its four-spinor can be split into two
independent Dirac equations with two-spinors associated with either spin up
or spin down. Each one of these two-spinors has upper and down components
associated with particle and antiparticle respectively. The absence of
angular momentum and spin-orbit interaction in the two-dimensional Dirac
equation as well as the use of 2$\times $2 matrices, instead of 4$\times $4
matrices, allow us to explore the physical consequences of the
negative-energy states in a mathematically simpler and more physically
transparent way.

The anomalous magnetic-like (tensor) coupling describes the interaction of
neutral fermions with electric fields and the bound states of fermions in
one-plus-one dimensions by a pseudoscalar double-step potential \cite{asc2}
and their scattering by a pseudoscalar step potential \cite{asc3} have
already been analyzed in the literature providing the opportunity to find
some quite interesting results. Indeed, the two-dimensional version of the
anomalous magnetic-like interaction linear in the radial coordinate,
christened by Moshinsky and Szczepaniak \cite{ms} as Dirac oscillator and
extensively studied before \cite{ito}-\cite{hug}, has also received
attention. Nogami and Toyama \cite{nt}, Toyama et al. \cite{tplus} and
Toyama and Nogami \cite{tn} studied the behaviour of wave packets under the
influence of that parity-conserving potential whereas Szmytkowski and
Gruchowski \cite{sg} proved the completeness of the eigenfunctions. More
recently Pacheco et al. \cite{pa} studied a few thermodynamics properties of
the 1+1 dimensional Dirac oscillator, and a generalization of the Dirac
oscillator for a negative coupling constant was presented in Ref. \cite{asc}%
. The two-dimensional generalized Dirac oscillator plus an inversely linear
potential has also been addressed \cite{asc4}. Furthermore, the
two-dimensional generalized Dirac oscillator plus scalar and vector harmonic
potentials have found a few applications relating nuclear phenomena \cite%
{PRC}.

The parity-conserving pseudoscalar potential $\sim \mathrm{tanh}\,\gamma x$
is of interest in quantum field theory where topological classical
backgrounds are responsible for inducing a fractional fermion number on the
vacuum. Models of these kinds, known as kink models are obtained in quantum
field theory as the continuum limit of linear polymer models \cite{gol}-\cite%
{sem}. Recently the complete set of bound states of fermions immersed in the
background of the pseudoscalar potential $V=\hbar c\gamma g\,\mathrm{tanh}%
\,\gamma x$, termed kink-like potential, has been addressed \cite{hott}.

In the present work the pseudoscalar potential $\sim \mathrm{tan}\,\gamma x$
is investigated. This trigonometric potential has a kink profile in a finite
region of space and reveals to be essentially confining. Beyond the
confinement property, this potential presents the harmonic oscillator and
the infinite square well as limit cases. A peculiar feature of this
potential, and for the potential analyzed in Ref. \cite{hott} as well, is
the absence of bound states in a nonrelativistic theory because it gives
rise to an ubiquitous repulsive potential. Fortunately, apart from solutions
corresponding to $|E|=mc^{2}$, the problem is reducible to the finite set of
solutions of the nonrelativistic exactly solvable symmetric  P\"{o}%
schl-Teller potential for both components of the Dirac spinor subject to a
constraint on their nodal structure. The whole spectrum of this
intrinsically relativistic problem is found analytically, if the fermion is
massless or not. A remarkable feature of this problem is the possibility of
trapping a fermion with an uncertainty in the position that can shrink
without limit as $|\gamma |$ and $|g|$ increase without violating the
Heisenberg uncertainty principle. This high degree of localization of
fermions in a single-particle interpretation of the theory is made plausible
by the introduction of the concept of effective wavelength.

\section{The Dirac equation with a pseudoscalar potential in a 1+1 dimension}

The 1+1 dimensional time-independent Dirac equation for a fermion of rest
mass $m$ coupled to a pseudoscalar potential reads

\begin{equation}
H\psi =E\psi ,\quad H=c\alpha p+\beta mc^{2}+\beta \gamma ^{5}V  \label{1}
\end{equation}

\noindent where $E$ is the energy of the fermion, $c$ is the velocity of
light and $p$ is the momentum operator. The positive definite function $%
|\psi |^{2}=\psi ^{\dagger }\psi $, satisfying a continuity equation, is
interpreted as a position probability density and its norm is a constant of
motion. This interpretation is completely satisfactory for single-particle
states \cite{tha}. We use $\alpha =\sigma _{1}$ and $\beta =\sigma _{3}$,
where $\sigma _{1}$ and $\sigma _{3}$ are Pauli matrices, and $\beta \gamma
^{5}=\sigma _{2}$. The charge conjugation operation requires that if $\psi $
is a solution with eigenenergy $E$ for the potential $V$ then $\sigma
_{1}\psi ^{\ast }$ is a solution with eigenenergy $-E$ for the potential $-V$%
. It is interesting to note that the unitary operation just exchanging the
upper and lower components of the Dirac spinor induced by $i\gamma ^{5}$
preserves the eigenenergies for a massless fermion when $V\rightarrow -V$.
The Dirac equation is covariant under $x\rightarrow -x$ if $V$ changes sign.
This is because the parity operator $P=\exp \left( \eta \right) P_{0}\sigma
_{3}$, where $\eta $ is a constant phase and $P_{0}$ changes $x$ into $-x$,
commutes with $\sigma _{3}$ but anticommutes with $\sigma _{1}$ and $\sigma
_{2}$.

Provided that the spinor is written in terms of the upper and the lower
components, $\psi _{+}$ and $\psi _{-}$ respectively, \noindent the Dirac
equation decomposes into:

\begin{equation}
\left( -E\pm mc^{2}\right) \psi _{\pm }=i\hbar c\psi _{\mp }^{\prime }\pm
iV\psi _{\mp }  \label{2}
\end{equation}

\noindent where the prime denotes differentiation with respect to $x$. In
terms of $\psi _{+}$ and $\psi _{-}$, defined on the closed interval $[a,b]$%
, the spinor is normalized as%
\begin{equation}
\int_{a}^{b}dx\left( |\psi _{+}|^{2}+|\psi _{-}|^{2}\right) =1  \label{2a}
\end{equation}%
\noindent so that $\psi _{+}$ and $\psi _{-}$ are square integrable
functions.

The boundary conditions on the eigenfunctions come into existence by
demanding that the Hamiltonian is Hermitian, viz.

\begin{equation}
\int_{a}^{b}dx\;\psi _{n}^{\dagger }\left( H\psi _{n^{\prime }}\right)
=\int_{a}^{b}dx\;\left( H\psi _{n}\right) ^{\dagger }\psi _{n^{\prime }}
\label{22-1}
\end{equation}

\noindent where $\psi _{n}$ is an eigenspinor corresponding to an eigenvalue
$E_{n}$. In passing, note that a necessary consequence of Eq. (\ref{22-1})
is that the eigenspinors corresponding to distinct effective eigenvalues are
orthogonal. It can be shown that (\ref{22-1}) is equivalent to

\begin{equation}
\left[ \psi _{n}^{\dagger }\sigma _{1}\psi _{n^{\prime }}\right]
_{x=a}^{x=b}=\left[ \left( \psi _{+}^{\ast }\right) _{n}\left( \psi
_{-}\right) _{n^{\prime }}+\left( \psi _{-}^{\ast }\right) _{n}\left( \psi
_{+}\right) _{n^{\prime }}\right] _{x=a}^{x=b}=0  \label{22-2}
\end{equation}%
It is clear from the pair of coupled first-order differential equations
given by (\ref{2}) that $\psi _{+}$ and $\psi _{-}$ have definite and
opposite parities if the pseudoscalar potential function is odd. In this
case, beyond the appropriate boundary conditions on the extremes of the
interval, we can impose boundary conditions at the origin in two distinct
ways: even functions obey the homogeneous Neumann condition ($d\psi
/dx|_{x=0}=0$) whereas odd functions obey the homogeneous Dirichlet
condition ($\psi \left( 0\right) =0$).

In the nonrelativistic approximation (potential energies small compared to $%
mc^{2}$ and $E\approx mc^{2}$) Eq. (\ref{2}) becomes

\begin{equation}
\psi _{-}=\left( \frac{p}{2mc}\,+i\,\frac{V}{2mc^{2}}\right) \psi _{+}
\label{3}
\end{equation}

\begin{equation}
\left( -\frac{\hbar ^{2}}{2m}\frac{d^{2}}{dx^{2}}+\frac{V^{2}}{2mc^{2}}+%
\frac{\hbar V^{\prime }}{2mc}\right) \psi _{+}=\left( E-mc^{2}\right) \psi
_{+}  \label{4}
\end{equation}

\noindent Eq. (\ref{3}) shows that $\psi _{-}$ is of order $v/c<<1$ relative
to $\psi _{+}$ and Eq. (\ref{4}) shows that $\psi _{+}$ obeys the Schr\"{o}%
dinger equation. Note that the pseudoscalar coupling results in the Schr\"{o}%
dinger equation with an effective potential in the nonrelativistic limit,
and not with the original potential itself. Indeed, this is the side effect
which in a 3+1 dimensional space-time makes the linear potential to manifest
itself as a harmonic oscillator plus a strong spin-orbit coupling in the
nonrelativistic regime \cite{ms}. The form in which the original potential
appears in the effective potential, the $V^{2}$ term, allows us to infer
that even a potential unbounded from below could be a binding potential.
This phenomenon is inconceivable if one starts with the original potential
in the nonrelativistic equation. \

It is also noticeable that the change $V\rightarrow V+$const into the Dirac
equation, and into its nonrelativistic limit as well, does not just implies
into the change $E\rightarrow E+$ const. Strange to say, the energy itself \
and not just the energy difference has physical significance. It has already
been verified that a constant added to the screened Coulomb potential \cite%
{asc10} \ as well as to the inversely linear potential \cite{asc20} is
undoubtedly physically relevant. As a matter of fact, it plays a crucial
role to ensure the existence of bound states in those cases.

For $E\neq \pm mc^{2}$, the coupling between the upper and the lower
components of the Dirac spinor can be formally eliminated when Eq. (\ref{2})
is written as second-order differential equations:

\begin{equation}
-\frac{\hbar ^{2}}{2}\,\psi _{\mp }^{\prime \prime }+\left( \frac{V^{2}}{%
2c^{2}}\mp \frac{\hbar }{2c}V^{\prime }\right) \,\psi _{\mp }=\frac{%
E^{2}-m^{2}c^{4}}{2c^{2}}\,\psi _{\mp }  \label{5}
\end{equation}

\noindent Here $V$ is the superpotential corresponding to the
Sturm-Liouville supersymmetric partner potentials \ $V^{2}/(2c^{2})\mp \hbar
V^{\prime }/(2c)$. \ This supersymmetric structure of the two-dimensional
Dirac equation with a pseudoscalar potential has already been appreciated in
the literature \cite{tn}, \cite{nog2} as has been too for a scalar potential
\cite{coop}. These last results show that the solution for this class of
problem consists in searching for bound-state solutions for two Schr\"{o}%
dinger equations. It should not be forgotten, though, that the equations for
$\psi _{+}$ or $\psi _{-}$ are not indeed independent because $E$ appears in
both equations. Therefore, one has to search for bound-state solutions for
both signals in (\ref{5}) with a common eigenvalue. At this stage one can
realize that \noindent \noindent the Dirac energy levels are symmetrical
about $E=0$. It means that the potential couples to the positive-energy
component of the spinor in the same way it couples to the negative-energy
component. In other words, this sort of potential couples to the mass of the
fermion instead of its charge so that there is no atmosphere for the
spontaneous production of particle-antiparticle pairs. Thus there is no room
for transitions from positive- to negative-energy solutions. This all means
that Klein\'{}s paradox never comes to the scenario.

The solutions for $E=\pm mc^{2}$, excluded from the Sturm-Liouville problem,
can be obtained directly from the Dirac equation (\ref{2}). One can observe
that such a sort of isolated solutions can be written as

\begin{eqnarray}
\psi _{\mp } &=&N_{\mp }\,\exp \left[ \mp v(x)\right]  \nonumber \\
&&  \label{30-3} \\
\psi _{\pm }^{\prime }\mp v^{\prime }\psi _{\pm } &=&\pm i\,\frac{2mc}{\hbar
}N_{\mp }\,\exp \left[ \mp v(x)\right]  \nonumber
\end{eqnarray}

\noindent where $N_{+}$ and $N_{-}$ are normalization constants and $%
v(x)=\int^{x}dy\,V(y)\,/(\hbar c)$. \noindent

The upper and the lower components can be normalized as $\int_{a}^{b}dx|\psi
_{\pm }|^{2}=|N_{\pm }|^{2}$ and the absolute values of the relative
normalization constants, $N_{+}$ and $N_{-}$, can be calculated from the
Dirac equation (\ref{2}). Indeed, one has%
\begin{eqnarray}
&&\left( E\pm mc^{2}\right) \int_{a}^{b}dx\,|\psi _{\mp }|^{2}=\left[ \left(
\hbar c\right) ^{2}\psi _{\pm }^{\ast }\psi _{\pm }^{\prime }\mp \hbar
cV|\psi _{\pm }|^{2}\right] _{x=a}^{x=b}  \nonumber \\
&&  \label{1000} \\
&&+2c^{2}\int_{a}^{b}dx\,\psi _{\pm }^{\ast }\left( -\frac{\hbar ^{2}}{2}%
\frac{d^{2}}{dx^{2}}\,+\frac{V^{2}}{2c^{2}}\pm \frac{\hbar }{2c}V^{\prime
}\right) \psi _{\pm }  \nonumber
\end{eqnarray}%
\noindent by imposing boundary conditions which do not break the condition
expressed by (\ref{22-2}), $\psi _{\pm }\left( b\right) =\psi _{\pm }\left(
a\right) =0$ for instance, the first term on the right-hand side of (\ref%
{1000}) vanishes. Hence one can conclude that
\begin{equation}
\int_{a}^{b}dx\,|\psi _{\pm }|^{2}=\frac{E\pm mc^{2}}{E\mp mc^{2}},\quad
\mathrm{for\quad }E\neq \pm mc^{2}  \label{1002}
\end{equation}%
\noindent and
\begin{equation}
-\frac{\hbar ^{2}}{2}\,\psi _{\mp }^{\prime \prime }+\left( \frac{V^{2}}{%
2c^{2}}\mp \frac{\hbar }{2c}V^{\prime }\right) \,\psi _{\mp }=0,\quad
\mathrm{for\quad }E=\pm mc^{2}  \label{1003}
\end{equation}%
\noindent Finally, use of (\ref{2a}) and (\ref{1002}) yields
\begin{equation}
N_{\pm }=\,\sqrt{\frac{E\pm mc^{2}}{2E}}  \label{22}
\end{equation}%
Of course, a possible solution with $E=+mc^{2}$ ($E=-mc^{2}$) has a Dirac
spinor with a vanishing lower (upper) component. One can observe that such
sort of isolated solution for $E=+mc^{2}$ is

\begin{equation}
\psi \sim \,e^{v}\left(
\begin{array}{l}
1 \\
0%
\end{array}%
\right)  \label{1a}
\end{equation}

\noindent and for $E=-mc^{2}$ is

\begin{equation}
\psi \sim \,e^{-v}\left(
\begin{array}{l}
0 \\
1%
\end{array}%
\right)  \label{1b}
\end{equation}

\noindent It is worthwhile to note that whereas one component of the Dirac
spinor vanishes the other one obeys a second-order differential equation
similar to (\ref{5}), viz.%
\begin{equation}
-\frac{\hbar ^{2}}{2}\,\psi _{\pm }^{\prime \prime }+\left( \frac{V^{2}}{%
2c^{2}}\pm \frac{\hbar }{2c}V^{\prime }\right) \,\psi _{\pm }=0\quad \mathrm{%
and}\quad \psi _{\mp }=0  \label{1c}
\end{equation}%
\noindent \noindent Of course well-behaved eigenstates are possible only if $%
V$ has an appropriate behaviour at the endpoints of the range $[a,b]$ \cite%
{hott}. It is noticeable that a possible solution with $E=-mc^{2}$
uncurtains a quintessentially relativistic solution.

\section{The trigonometric tangent potential}

Now let us concentrate our attention on the potential
\begin{equation}
V=\hbar c\gamma g\,\mathrm{tan}\,\gamma x  \label{7}
\end{equation}%
\noindent where the kink parameter, $\gamma $, and the dimensionless
coupling constant, $g$, are real numbers. Due to the infinities at $x=\pm
\pi /(2|\gamma |)$ we restrict ourselves to $|x|\leq \pi /(2|\gamma |)$.
This potential is unbounded from below so that it is unable to bind a
fermion in the nonrelativistic theory. The potential is invariant under the
change $\gamma \rightarrow -\gamma $ so that the results can depend only on $%
|\gamma |$ whereas the sign of $V$ depends on the sign of $g$. Since the
solutions for different signs of $g$ can be connected by the charge
conjugation transformation, and also by the chiral transformation in the
event of massless fermions, we restrict ourselves to the case $g>0$.

The Sturm-Liouville problem corresponding to Eq. (\ref{5}) becomes

\begin{equation}
-\frac{\hbar ^{2}}{2}\,\psi _{\pm }^{\prime \prime }+V_{\mathtt{eff}}^{[\pm
]}\,\psi _{\pm }=E_{\mathtt{eff}}\,\psi _{\pm }  \label{8}
\end{equation}

\noindent where we recognize the effective potential as the exactly solvable
symmetric P\"{o}schl-Teller potential \cite{pt}-\cite{nie}

\begin{equation}
V_{\mathtt{eff}}^{[\pm ]}(x)=\frac{\hbar ^{2}\gamma ^{2}}{2}\left[ g\left(
g\pm 1\right) \,\mathrm{tan}^{2}\,\gamma x\pm g\right]  \label{9}
\end{equation}%
whose normalizable eigenfunctions corresponding to bound-state solutions,
subject to the boundary conditions $\psi _{\pm }=0$ as $|x|=\pi /(2|\gamma
|) $ (where the potential becomes infinitely steep) and identically zero for
$|x|>\pi /(2|\gamma |)$, are possible only if the effective potentials for
both $\psi _{+}$ and $\psi _{-}$ present potential-well structures.
According to (\ref{9}), this demands that $g>1$. The corresponding effective
eigenenergy is given by (in the notation of Refs. \cite{nie1}-\cite{nie} $%
g\left( g\pm 1\right) =\lambda \left( \lambda -1\right) $)

\noindent

\begin{equation}
E_{\mathtt{eff}}=\frac{E^{2}-m^{2}c^{4}}{2c^{2}}=\,\frac{\hbar ^{2}\gamma
^{2}}{2}\left( n_{\pm }^{2}+2n_{\pm }\lambda _{\pm }+\lambda _{\pm }\pm
g\right)  \label{10}
\end{equation}

\noindent where

\begin{equation}
\lambda _{+}=g+1,\quad \lambda _{-}=g,\quad n_{\pm }=0,1,2,\ldots \quad
\label{20}
\end{equation}

\noindent Notice that $V_{\mathtt{eff}}^{[\pm ]}$ is an even function under $%
x\rightarrow -x$. Furthermore, the capacity of the potential to hold
bound-state solutions is infinite. In fact, the effective potential is a
well potential limited by infinite barriers at $x=\pi /(2|\gamma |)$.
Referring to (\ref{9}) and (\ref{10}) one can note that the Dirac
eigenenergies are restricted to the range
\begin{equation}
|E|>\sqrt{m^{2}c^{4}+\left( \hbar c\gamma \right) ^{2}g}  \label{100}
\end{equation}%
and that there is no continuum. Since the positive- and
negative-eigenenergies never intercept once again one can see that Klein\'{}%
s paradox is absent from the scenario. In other words, the pseudoscalar
tangent potential is a true confining potential. Furthermore, the fermion
tends to be confined into a region of space which tends to zero as $|\gamma
|\rightarrow \infty $. \ In order to match the common effective eigenvalue
for the effective potentials $V_{\mathtt{eff}}^{[+]}$ and $V_{\mathtt{eff}%
}^{[-]}$, one can see from (\ref{10})-(\ref{20}) that there appears the
constraint
\begin{equation}
n_{-}=n_{+}+1  \label{10f}
\end{equation}%
requiring that $n_{-}=1,2,3,\ldots $ This last fact can be better understood
by observing that $V_{\mathtt{eff}}^{[-]}$ is deeper than $V_{\mathtt{eff}%
}^{[+]}$. Now, (\ref{10})-(\ref{20}) tell us that

\begin{equation}
E=\pm \,\sqrt{m^{2}c^{4}+\left( \hbar c\gamma \right) ^{2}\left[
n_{+}^{2}+2n_{+}\left( g+1\right) +2g+1\right] }  \label{10e}
\end{equation}%
The upper and lower components of the Dirac spinor can be written as (see
Ref. \cite{nie1}-\cite{nie})%
\begin{equation}
\psi _{\pm }=N_{\pm }\,\sqrt{|\gamma |\left( n_{\pm }+\lambda _{\pm }\right)
\frac{\Gamma \left( 2\lambda _{\pm }+n_{_{\pm }}\right) }{\Gamma \left(
n_{_{\pm }}+1\right) }}\,\left( 1-z^{2}\right) ^{1/4}P_{n_{_{\pm }}+\lambda
_{\pm }-1/2}^{1/2-\lambda _{\pm }}\left( z\right)  \label{21}
\end{equation}

\noindent where $z=\sin \gamma x$ and $P_{\nu }^{\mu }\left( z\right) $ is
the associated Legendre function of the first kind. In terms of the
Gegenbauer (ultraspherical) polynomial, $C_{n}^{\left( a\right) }\left(
z\right) $, a polynomial of degree $n$ defined on the interval $[-1,+1]$,
the components of the Dirac spinor can be written as%
\begin{equation}
\psi _{\pm }=N_{\pm }\,2^{-\lambda _{\pm }}\sqrt{2|\gamma |\left( n_{\pm
}+\lambda _{\pm }\right) \frac{\Gamma \left( n_{_{\pm }}+1\right) }{\Gamma
\left( n_{_{\pm }}+2\lambda _{\pm }\right) }}\,\frac{\Gamma \left( 2\lambda
_{\pm }\right) }{\Gamma \left( \lambda _{\pm }+1/2\right) }\left(
1-z^{2}\right) ^{\lambda _{\pm }/2}C_{n_{_{\pm }}}^{\left( \lambda _{\pm
}\right) }\left( z\right)  \label{210}
\end{equation}%
Since $C_{n}^{\left( a\right) }\left( -z\right) =\left( -\right)
^{n}C_{n}^{\left( a\right) }\left( z\right) $ and $C_{n}^{\left( a\right)
}\left( z\right) $ has $n$ distinct zeros (see, e.g. \cite{abr}), it becomes
clear that $\psi _{+}$ and $\psi _{-}$ have definite and opposite parities,
as expected. Furthermore, the number of nodes of $\psi _{+}$ and $\psi _{-}$
just differ by $\pm 1$ according to the rule expressed by (\ref{10f}). Note
that these solutions for the second-order differential equations given by (%
\ref{8}), for $E\neq -mc^{2}$, is entirely equivalent to the Dirac equation
itself provided $N_{\pm }$ satisfy Eq. (22).

It is noteworthy that the width of the position probability density
decreases as $|\gamma |$ or $g$ increases. As such it promises that the
uncertainty in the position can shrink without limit. It seems that the
uncertainty principle dies away provided such a principle implies that it is
impossible to localize a particle into a region of space less than half of
its Compton wavelength (see, for example, \cite{str}). This apparent
contradiction can be remedied by resorting to the concept of effective mass.
The previous results suggest that one can define the effective mass as%
\begin{equation}
m_{\mathtt{eff}}=\sqrt{m^{2}+\left( \frac{\hbar \gamma }{c}\right) ^{2}g}
\label{400}
\end{equation}%
in such a way that the Dirac eigenenergies are restricted to the range $%
|E|>m_{\mathtt{eff}}c^{2}$. Now it is possible to define the effective
Compton wavelength as $\lambda _{\mathtt{eff}}=\hbar /(m_{\mathtt{eff}}c)$.
Hence, the minimum uncertainty in the position consonant with the
uncertainty principle is given by $\lambda _{\mathtt{eff}}/2$ whereas the
maximum uncertainty in the momentum is given by $m_{\mathtt{eff}}c$. It
means that the localization of a neutral fermion can shrink to zero without
spoil the single-particle interpretation of the Dirac equation, even if the
trapped neutral fermion is massless. It is true that as $|\gamma |$ or $g$
increases the binding potential becomes stronger, though, it contributes to
increase the effective mass of the fermion in such a way that there is no
energy available to produce fermion-antifermion pairs.

Turning now to the isolated solutions, one can observe from (\ref{1a}) and (%
\ref{1b}) that a normalizable isolated solution is possible only if the
upper component of the spinor vanishes and $E=-mc^{2}$. The normalized Dirac
spinor can be written as
\begin{equation}
\psi =2^{-g+1/2}\sqrt{|\gamma |g\,\frac{\Gamma \left( 2g\right) }{\Gamma
\left( g+1/2\right) }}\left( 1-z^{2}\right) ^{g/2}\left(
\begin{array}{l}
0 \\
1%
\end{array}%
\right)  \label{7a}
\end{equation}%
independently of the magnitude of $g$. It turns out that $g>1$. Indeed, $%
\psi _{-}$ satisfies Eq. (\ref{8}) with $V_{\mathtt{eff}}^{[-]}$ given by (%
\ref{9}) ($E_{\mathtt{eff}}=0$). Therefore, the coupling constant for an
isolated solution has precisely the same restriction as that one for the
solutions of the Sturm-Liouville problem. After all, the best localization
of fermions as well as the validity of the uncertainty principle is
unperturbed if one uses the effective mass given by (\ref{400}). As a matter
of fact, a numerical calculation for the most critical case ($m=0$) with $%
g=1.001$ yields $\Delta x=0.5680\lambda _{\mathtt{eff}}$ $\ $and $\Delta
p=0.9995m_{\mathtt{eff}}c$, regardless the value of $\gamma $ ($\hbar =c=1$%
).\noindent

\section{Conclusions}

We have succeeded in searching for the complete set of exact bound-state
solutions of fermions in the background of a pseudoscalar trigonometric
tangent potential. This kind of potential has opposite values at the ends of
the interval, viz. $V\left( +\pi /(2|\gamma |)\right) =-V\left( -\pi
/(2|\gamma |)\right) $. It is this topological behaviour that gives rise to
two different kinds of solutions. The potential admits no scattering states
and, except for the solution $E=-mc^{2}$, it presents a spectral gap greater
than $2m_{\mathtt{eff}}c^{2}$. Since $C_{0}^{\left( a\right) }\left(
z\right) =1$ (see, e.g. \cite{abr}) and $N_{+}=0$ for $E=-mc^{2}$, one can
see that the position probability amplitude corresponding to the isolated
solution given by (\ref{7a}) can be written in the very same mathematical
structure of the remaining amplitudes. Thus, one could suspect that the
isolated solution is just a particular case and that this segregation is a
particularity of the method used in this paper. However, the isolated
solution has some distinctive characteristics when compared to the solutions
of the Sturm-Liouville problem which lead us to believe that, in fact, they
belong to different classes of solutions. The isolated solution breaks the
symmetry of the energy levels about $E=0$ exhibited by the solutions of the
Sturm-Liouville problem, and the corresponding eigenspinor has one component
differing from zero. It is this asymmetric spectral behaviour that leads to
the fractionalization of the fermion number in quantum field theory \cite%
{sem}.

For massless fermions, except for $E=0$, the spectral gap is greater than $2%
\sqrt{3}\hbar c|\gamma |$ and the Dirac Hamiltonian anticommutes with $%
\sigma _{3}$ in such a way that the positive- and negative-eigenenergy
solutions can be mapped by the operation $\psi _{-E}=\sigma _{3}\psi _{E}$.
The solution given by (\ref{7a}) appears now in the center of the spectral
gap.

As mentioned in the Introduction of this work, the anomalous magnetic-like
coupling turns into a pseudoscalar coupling when the fermion experiences an
one-dimensional motion. The anomalous magnetic interaction has the form $%
-i\mu \beta \vec{\alpha}\cdot \vec{\triangledown}\phi (r)$, where $\mu $ is
the anomalous magnetic moment in units of the Bohr magneton and $\phi $ is
the electric potential, i.e., the time component of a vector potential \cite%
{tha}. In one-plus-one dimensions the anomalous magnetic interaction turns
into $\sigma _{2}\mu \phi ^{\prime }$, then one might suppose that the
trigonometric tangent potential is due to an electric potential proportional
to $\ln \left( \mathrm{cos}\,\gamma x\right) ^{g}$. Therefore, the problem
addressed in this paper could be considered as that one of confining neutral
fermions by a bowl-shaped electric potential.

\bigskip \bigskip \bigskip

\noindent \textbf{Acknowledgments}

The authors wish to thank Prof. M. Hott for stimulating discussions
and for the algebraic calculations leading to Eq. (\ref{22})and to
an anonymous referee for very constructive remarks.  This work was
supported in part by means of funds provided by CAPES, CNPq and
FAPESP.

\end{document}